\documentclass{article}
\usepackage[labelformat=simple]{subcaption}

\DeclareCaptionLabelFormat{subcaptionlabel}{\normalfont(\textbf{#2}\normalfont)}

\usepackage[english]{babel}

\usepackage[letterpaper,top=2cm,bottom=2cm,left=3cm,right=3cm,marginparwidth=1.75cm]{geometry}

\usepackage{amsmath}
\usepackage{lscape}
\usepackage{graphicx}
\usepackage[colorlinks=true, allcolors=blue]{hyperref}
\usepackage{natbib}

\title{On the Galactic Halos Rotation by \textit{Planck} Data}

\author{Noraiz Tahir $^{1,2,*}$, Francesco De Paolis $^{1,2,3}$, Asghar Qadir $^{4}$, \\ Achille A. Nucita $^{1,2,3}$\\
$^1$ Department of Mathematics and Physics ``Ennio De Giorgi'',
University of Salento, \\ Via per Arnesano, I-73100  Lecce, Italy;\\ francesco.depaolis@le.infn.it (F.D.P.); achille.nucita@le.infn.it (A.A.N.)\\
$^2$ INFN, Sezione di Lecce, Via per Arnesano, I-73100 Lecce, Italy\\
$^3$ INAF, Sezione di Lecce, Via per Arnesano, I-73100 Lecce, Italy  \\
$^4$ Abdul Salam School of Mathematics, G. C. University, Lahore 54600, Pakistan; \\ asgharqadir46@gmail.com (A.Q.)\\
}

\begin{document}
\maketitle

\begin{abstract}
As galactic halos are not directly visible, there are many ambiguities regarding their composition and rotational velocity. Though most of the dark matter is non-baryonic, {\it some fraction is}, and it can be used to trace the halo rotation. Asymmetries in the CMB towards M31 had been seen in the Planck data and ascribed to the rotational Doppler shift of the M31 halo. Subsequently, the same methods were used in the direction of five other galaxies belonging to the Local Group, namely M33, M81, M82, NGC 5128, and NGC 4594. It had been proved that there could be stable clouds of gas and dust in thermal equilibrium with the CMB at 2.7 K, which had been called ``virial clouds''. In this paper, adopting this scenerio, an attempt is made to constrain the fraction of dust grains and gas molecules in the clouds.
\end{abstract}


\section{Introduction}
According to the standard cosmological model ($\Lambda$CDM) baryons contribute $\sim$5\% of the Universe, of which we have only seen about half \citep{bertone_2010, burdyuzha2020, perez2020}. Where the rest of them are and how they can be detected are still open questions. This is the so-called ``missing baryon problem'' (see e.g., \cite{shull2012missingbaryons, driver2021missingbaryons}). It has been suggested that the missing baryons might have been ejected from the galaxies into the intergalactic medium (IGM), and~a {\it fraction} of that component may be present in the warm-hot medium around galaxies, at~temperatures of about $10^5$--$10^7$ K (for details see Refs. \citep{cen1999baryons, cen2006baryons, fraser2011estimate, gupta2012huge}). On the other hand, it is also suggested that a non-negligible fraction of these baryons are in the form of ``cold gas clouds'' spread out in the galactic halos \citep{de1995case, gerhard1995baryonic, pifenniger1994baryonic, palla1983firststars}.

In 1995, it was proposed that there may be molecular hydrogen (${\rm H_2}$) clouds in galactic halos at exactly the CMB temperature, so that they do not stand out against it \citep{de1995scenario}. It had been assumed, without suggesting any model or mechanism for it, that the clouds remain stable and avoid further gravitational collapse. It has been argued that cold clouds {\it could not} be in equilibrium with the CMB, as~they would not be able to absorb the long wavelengths in it \citep{alcock1993possible,de1991nature}, and~so would have to collapse. To see such clouds merged in the CMB, in~galactic halos, it was suggested that one look for $\gamma$-ray scintillation due to cosmic rays striking the ${\rm H_2}$ molecules in the clouds. Observations by EGRET did display an excess of diffuse $\gamma$-ray emission in the halo of the Milky Way, but it could also have been due to unresolved sources in the Galaxy or from high-latitude inverse Compton scattering (see \citep{dixon1998evidence, de1998gamma,1999AA}). Another suggestion for detecting the claimed gas clouds is
to look for an asymmetric Doppler shift due to the rotation of the galaxies. The suggestion had been to consider our sister galaxy Andromeda (M31), and~assume that the clouds rotate with the whole M31 halo. It is clear that the rotation would induce a red-shift or a blue-shift of the emitted electromagnetic signal depending on the rotation direction with respect to the observer \citep{de1995observing}.

In a series of papers, first with the analysis of WMAP and then of {\it Planck} data, CMB data was used to trace the halos of some nearby galaxies \citep{de2011possible, de2014planck, de2015planck, gurzadyan2015planck, de2016triangulum, gurzadyan2018messier, de2019rotating} to determine whether if the data supports the Doppler shift explanation of the temperature asymmetry. The data uncovered it, not only for the galactic disks but also for some galactic halos. Further support for the Doppler shift came from the observation that the temperature asymmetry was almost frequency-independent.

In summary, there is no doubt that this phenomenon is seen in several nearby galaxies. It remains to investigate whether it is related to cold halo gas clouds, with or without dust contamination, or is due to other effects. As a working hypothesis here, we assume that it is caused, at~least in part, by virial clouds populating the galactic halos. For this purpose, we model these clouds and estimate their physical parameters.

This paper aims to present a model to see the change in the physical parameters of virial clouds when they are contaminated by different percentages of dust grains. Then we will try to estimate the rotational velocity of the halos of several nearby edge-on spiral galaxies, which show a consistent temperature asymmetry in the CMB data as observed by the {\it Planck} satellite and already analyzed. The gas and dust clouds had been modeled previously in Refs.~\cite{tahir2019constraining, tahir2019seeing} by using ad-hoc boundary conditions and then the halo rotational velocities of the above-mentioned spiral galaxies estimated. We now use a more reliable model described in detail in Ref.~\cite{qadir2019virial} to estimate the physical parameters of virial clouds with contamination of halo dust grains in them.

The plan of the paper is as follows: in Section~\ref{three}, the physical parameters of virial clouds are estimated, as~a function of the fraction of molecular hydrogen (${\rm H_2}$), helium (He) and interstellar dust. In~Section~\ref{model}, the adopted models describing the distribution of the gas clouds in the halo of the considered galaxies (M31, M33, M81, M82, NGC 5128, and~NGC 4594)  are briefly described. These models are those generally adopted for describing the halo dark matter distribution and include the Navarro--Frenk--White (NFW), Moore, and Burkert models. Using these models, we then derive the distribution of the virial clouds and the cloud-filling factor. Then, in~Section~\ref{rotation}, we give an expression to estimate the galactic halo rotational velocity. In~Section~\ref{halomodel}, the halo rotational velocity is estimated, assuming that each virial cloud is optically thick and the whole halo is composed of such clouds.  However, the~results obtained can be easily generalized to the case in which the virial clouds constitute only a fraction of the galactic halo dark matter and, even if less straightforwardly, to~the not-optically thick clouds. Finally, in~Section~\ref{results}, the~obtained results are~discussed.

\section{\label{three} Virial Cloud~Model}
Virial clouds populating the galactic halos are assumed to be spherically symmetric and at a temperature equal to that of the CMB, i.e.,~$T\simeq 2.7$ K. Each cloud is assumed to be composed of molecular hydrogen, helium, and~interstellar dust grains with different fractions. It has been suggested that galactic halos contain interstellar dust grain fraction of about 1--3$\%$ in mass, and~the rest is ${\rm H_2}$, He and other heavier molecules \citep{imara2018model, draine2003interstellar}. Halo dust grains are expected to be at a slightly higher temperature with respect to that of the gas, about 5 K or so, depending on the cloud position within the galactic halo \citep{yershov2020distant}, but~as a first approximation, we assume here that dust and gas are at the same temperature, i.e.,~$2.7$ K.

There is no doubt that interstellar dust grains are found in galactic halos. It can be easily expected that dust grains form in the disk, and then the radiation pressure from disk stars, exerting a force on the dust grains, can push them towards the external regions of galaxies \citet{ferrara1991evolution}. This force can move the dust grains to high galactic latitudes via dust-gas coupling. This process is similar to the driving agent of cool winds in giant stars  \citep{1974ApJ...193..585S, 1975ApJ...198..583K}. Once these grains are above the disk, i.e.,~in the galactic halo, the the~radiation pressure can expel the grains out of the galaxy environment \citep{chiao1972radiation, greenberg1987there, barsella1989large}. The important question that needs to be answered is then to  quantitatively estimate how much these dust grains contaminate the galactic halos.

Now, an objection possibly arising is whether magnetic fields and the particle destruction rate in the hot halos can play a major role in the dust grain evolution so that, eventually, these grains might not move outside the hot halo environment. In this respect, \citet{ferrara1991evolution} have shown that different dust grain compounds respond differently to ambient conditions. For example, graphites move faster than silicates, since graphite is much lighter than silicates grains, but both types of grains can reach values in excess of about 100 km s$^{-1}$; thus, they can move further outside the halos depending upon the halo gas temperature, galaxy mass, and~radiation pressure coefficients. It is also found in the literature that most of the halo dust grains are in the form of amorphous carbon, and~silicates, all the other kinds of molecules and atoms that are thought to be present might be dominated by the effects of the hot gaseous halos, i.e., the high radiation pressure and magnetic field. Hence, the molecules and atoms other than graphite and silicates might not be able to reach the outer parts of the halos. So, these outer regions of the halos might contain traces of amorphous carbon and silicate dust grains, whose masses are quite similar. Therefore, the average mass of a single halo dust grain can be assumed to be $\simeq 5.65 \times 10^{-23}$ g \citep{hirashita2020dust}.

The cloud  mass within the galactocentric distance $r$ can be written as
\begin{equation}
M_{cl}(r)=\alpha M_{H_2}(r)+\beta M_{He}(r)+ \gamma M_{d}(r),
\label{three01}
\end{equation}
where $\alpha$, $\beta$, and~$\gamma$ are the fractions of $H_{2}$, He, and dust, with~the condition that $\alpha+\beta+\gamma=1$. Of~course, $M_{H_2}(r)$, $M_{He}(r)$, and~$M_{d}(r)$ are the total mass of ${\rm H_{2}}$, He and dust within the distance $r$. If~the cloud is at the CMB temperature; hence, the~CMB can be considered a heat bath for the virial clouds. One can use the canonical ensemble distribution to obtain the density profile of these clouds, which can be written as
\begin{eqnarray}
\rho_{cl}^3(r)=\left(\frac{512\pi^{9/2}}{3^{9/2}}\right)\left(\frac{G}{kT}\right)^{9/2} (\rho_{c_{H_2}}\rho_{c_{He}}\rho_{c_d})^{3/2} \times (m_{H_2}m_{He}m_d)^{5/2} \delta,
\label{three02}
\end{eqnarray}
where
\begin{eqnarray}
\delta= exp\left[-\left(A+B+C\right)\right].
\label{three03}
\end{eqnarray}

Here,
\begin{equation}
A=\frac{\alpha GM_{H_2}(r)m_{H_2}}{rkT},
\label{three04}
\end{equation}
\begin{equation}
B=\frac{\beta GM_{He}(r)m_{He}}{rkT},
\label{three05}
\end{equation}
and
\begin{equation}
C=\frac{\gamma GM_d(r)m_d}{rkT}.
\label{three06}
\end{equation}

Here, $G$ is Newton's gravitational constant, $k$ is the Boltzmann constant, $T$ is the cloud temperature, $\rho_{c_{H_2}}$, $\rho_{c_{He}}$, and $\rho_{c_d}$ are the central density of ${\rm H_2}$, He and dust, $m_{H_2}$, $m_{He}$, and~$m_{d}$ are the masses of the single ${\rm H_2}$ molecule, He atom, and halo dust grain, respectively. Taking the natural logarithm of Equation~(\ref{three02}), and~then substituting the obtained expression in Equation~(\ref{three01}), we obtain
\begin{align}
\alpha M_{H_2}(r)m_{H_2}&+\beta M_{He}(r)m_{He}+\gamma M_{d}(r)m_d= -\frac{3rkT}{4\pi G} \ln \left(\frac{\rho_{cl}(r)}{\lambda}\right),
\label{threen02}
\end{align}
where $\lambda=(8\pi^{3/2}/3^{3/2})[G/(kT)]^{3/2}(\rho_{c_{H_2}}\rho_{c_{He}}\rho_{c_d})^{1/2} [m_{H_2}m_{He}m_d]^{5/6}$. The~boundary conditions are at $r=0$, where $\rho_{cl}\rightarrow\rho_c$, and~$(d\rho(r)/dr)|_{r\rightarrow0}=0$. Taking the derivative of Equation~(\ref{three02}) with respect to $r$ and using Equation~(\ref{threen02}), we obtain the following differential equation
\begin{align}
r\frac{d\rho_{cl}(r)}{dr}-r^2\left(\frac{2\pi G}{kT}\right)[\rho_{cl}(r)(& \alpha\rho_{c_{H_2}} m_{H_2} + \beta\rho_{c_{He}}m_{He} +\gamma \rho_{c_d}m_d)] \nonumber \\  &-\rho_{cl}(r)\ln\left(\frac{\rho_{cl}(r)}{\lambda}\right)=0,
\label{three07}
\end{align}

\begin{figure}[ht]
\centering
\includegraphics[width=10.5 cm]{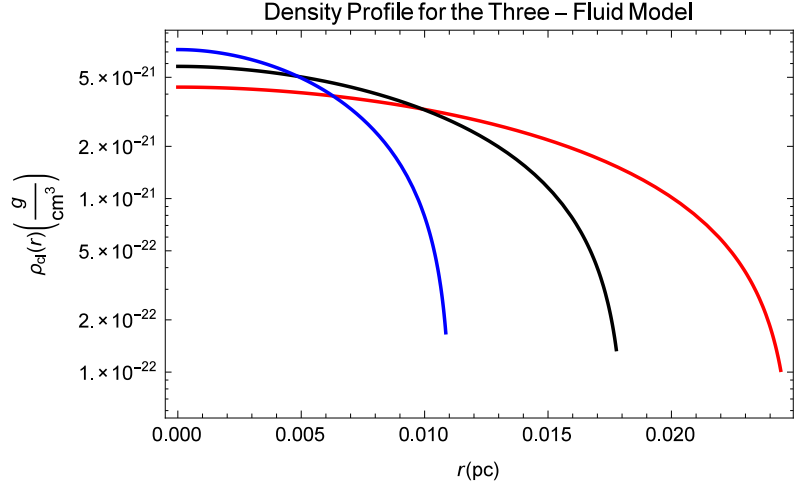}
\caption{The density profile of three fluids virial cloud with different fractions of ${\rm H_2}$, He, and~dust. The~red curve is for parameter values $\alpha=0.74$, $\beta=0.25$, and $\gamma=0.01$; the black curve corresponds to the profile for  $\alpha=0.74$, $\beta=0.24$, and $\gamma=0.02$; the blue curve is obtained in the case of  $\alpha=0.73$, $\beta=0.24$, and $\gamma=0.03$.\label{figthreefluid}}
\end{figure}

In Figure~\ref{figthreefluid}, we give the density profile of virial clouds with different fractions of dust, i.e.,~$\gamma=0.01$ (red curve), $0.02$ (black curve), and~$0.03$ (blue curve). As one can see, by increasing the contamination of dust grains, the cloud's central density remains approximately the same, but the size of the cloud decreases, and so as its mass. This occurs because heavier molecules are pulled in more with gravity, so the cloud shrinks in size. In~Table~\ref{tab1}, we give the physical parameters (cloud central density, Jeans mass, and~radius) of the three fluids virial cloud with different values of the parameters $\alpha$, $\beta$, and~$\gamma$.

\begin{table}
\centering
\begin{tabular}{ |c|c|c|c|c|c| } 
\hline
\boldmath{$\alpha$} & \boldmath{$\beta$} & \boldmath{$\gamma$} & \boldmath{$\rho_c$} & \boldmath{$M_J$} & \boldmath{$R_J$} \\
			&&& \boldmath{$(10^{-21} \, {\rm g \,cm^{-3}})$} & \boldmath{$M_{\odot}$}& \textbf{pc} \\
\hline
0.74 & 0.01 & 0.25 & 1.44 & 0.020 & 0.025\\
0.74 & 0.02 & 0.24 & 1.79 & 0.019 & 0.018\\
0.73 & 0.03 & 0.24 & 2.04 & 0.017 & 0.011\\
\hline
\end{tabular}
\caption{Physical parameters of three fluids virial cloud. The~virial cloud central  density, Jeans mass, and~radius are given for three different choices of the parameters $\alpha$, $\beta$, and~$\gamma$, which define the cloud composition (see text for details). \label{tab1}}
\end{table}
\section{\label{model} Virial clouds distribution in the galactic~halos}
As next, we assume that a fraction $f$ of the galactic halo dark matter is in the form of virial clouds. If these clouds are contributing to the halo dark matter, we can say that their distribution  follows the dark matter distribution profile, which is generally modeled by using the Navarro--Frenk--White (NFW), the~Moore, and~the Burkert density profiles \cite{1996ApJ...462..563N, 1999ApJ...524L..19M, 1995ApJ...447L..25B}.
These radial density profiles are, respectively,
\begin{equation}
{\displaystyle \rho_{_{N}}(r)=\frac{\rho_{\circ}^{N}}{\left(\frac{r}{r_{\circ}^{N}}\right)\left(1+\frac{r}{r_{\circ}^{N}}\right)^{2}}},
\label{eq1}
\end{equation}
(NFW model)
\begin{equation}
{\displaystyle \rho_{_M}(r)=\frac{\rho_{\circ}^{M}}{\left(\frac{r}{r_{\circ}^{M}}\right)^{1.5}\left(1+\left(\frac{r}{r_{\circ}^{M}}\right)^{1.5}\right)}},
\label{eq2}
\end{equation}
(Moore model)

and
\begin{equation}
{\displaystyle \rho_{_B}(r)=\frac{\rho_{\circ}^{B}}{\left(1+\frac{r}{r_{\circ}^{B}}\right)\left(1+\left(\frac{r}{r_{\circ}^{B}}\right)^{2}\right)}}.
\label{eq3}
\end{equation}
(Burkert model).

Here, $r_{\circ}^{N,M,B}$ are the halo core radius of the adopted models, and $\rho_{\circ}^{N,M,B}$ are the relative core densities. It can be easily seen that the core density corresponds to the central density of the considered galactic model. Of course, the values of both the core radius and core density depend on the considered halo model and are given in columns 3--8 of Table~\ref{tab2}. These values are taken from \cite{tamm2012stellar, tahir2019constraining} and are reproduced  in Table~\ref{tab2} for the advantage of the reader.

\begin{table}
\centering
\begin{tabular}{ |c|c|c|c|c|c|c|c| } 
\hline
\textbf{Galaxy} &\boldmath{$i$} &\boldmath{$r_{\circ}^N$} &\boldmath{$r_{\circ}^M$} &\boldmath{$r_{\circ}^B$} &\boldmath{$\rho_{\circ}^N$} &\textbf{ $\rho_{\circ}^M$} &\boldmath{$\rho_{\circ}^B$} \\
  & \textbf{(deg)} & \textbf{kpc} & \textbf{kpc} & \textbf{kpc} &\boldmath{$({\rm g~cm^{-3}})$} &\boldmath{$({\rm g~cm^{-3}})$ }&\boldmath{$ ({\rm g~cm^{-3}})$} \\
\hline
M31 & 77 &16.5 & 31 & 9.1 & $3.06\times 10^{-24}$ & $5.05\times 10^{-24}$ & $1.25\times 10^{-23}$ \\
M33 & 59 &35 & 18 & 12 & $7.00\times 10^{-25}$ & $7.60 \times 10^{-26}$ & $5.00\times 10^{-25}$ \\
M81 & 14 &10 & 18 & 8.9 & $5.10\times 10^{-24}$ & $7.65\times 10^{-25}$ & $3.68 \times 10^{-24}$ \\
M82 & 26 &11 & 13 & 9 &  $4.00 \times 10^{-27}$ & $1.79\times 10^{-27}$ & $6.29\times 10^{-27}$ \\
NGC 5128 & 14.6 & 21 & 12 & 9.8 &$1.67\times 10^{-25}$ & $4.40\times 10^{-25}$ & $1.00\times 10^{-24}$
\\
NGC 4594 & 83 & 4.5 & 7.4 & 3.2 &$1.96\times 10^{-22}$ & $3.18\times 10^{-23}$ & $3.93\times 10^{-22}$
\\
\hline
\end{tabular}
\caption{Adopted physical parameters for the NFW, Moore, and Burkert models for each considered galaxy. Note: For  each considered galaxy, indicated  in column 1, the inclination angle $i$ is given in the second column, the core radius for each adopted model is also given (columns 3--5), together with the respective  core density (columns 6--8).	\label{tab2}}
\end{table}

The surface density of the virial clouds obtained in a galactic halo is given by \citep{lokas2001properties}
\begin{equation}
	\Sigma(a)=2\int_{a}^{\infty} \frac{r \rho_{_{N,M,B}}(r)}{(r^2-a^2)^{1/2}} dr.
	\label{d1}
\end{equation}
and, therefore, the total number of virial clouds per unit area can be estimated as
\begin{equation}
{\displaystyle N(a)=f\left(\frac{\Sigma(a)}{M_J}\right)},
\label{eq4}
\end{equation}
where $f$ is the fraction of halo dark matter present in the form of virial clouds, $M_J$ is the Jeans mass of a single virial cloud, and~$a$ is a given value of the halo projected~radius.

Once we have estimated the total number of virial clouds in the three considered galactic dark matter halo models using Equation~(\ref{eq4}), we are in the position to estimate the cloud filling factor $S$, which is actually the ratio of the area covered by all the clouds within the halo radius $R$ and the total projected area and is given by
\begin{equation}
{\displaystyle S(\leq R)=\frac{\int_{0}^{R}2 \pi a N(a) r_{cl}^2 da}{\int_{0}^{R}2\pi a da}},
\label{eq5}
\end{equation}
where $r_{cl}$ is the virial cloud radius, and~$M_{cl}$ its~mass.

\section{\label{rotation} Estimating the galaxy halo rotational~velocity}
Starting from the cumulative radial distribution of the virial clouds and considering the temperature asymmetry in the CMB data towards a given galaxy, we can then estimate the halo rotational velocity, which is given by
\begin{equation}
{\rm v_{rot}}=\left(\frac{\Delta T}{T}\right)\left(\frac{c}{2\sin(i)Sf<\tau>}\right),
\label{eq6}
\end{equation}
where $c$ is the light speed in a vacuum, $i$ is the inclination angle of the considered galaxy, $S$ is the virial cloud filling factor as calculated from Equation~(\ref{eq5}), $<\tau>$ is the average optical depth of the virial clouds,  and~$\Delta T/T$ is the temperature asymmetry detected in the CMB towards the two opposite quadrants of the galactic halo (for details see \citep{de2014planck, de2015planck, gurzadyan2015planck, de2016triangulum, gurzadyan2018messier, de2019rotating}). The~values of $\Delta T/T$ are obtained from the {\it Planck} data analyzed for the SMICA band and previously analyzed in Refs.~\cite{de2016triangulum, gurzadyan2018messier, tahir2019constraining}. For~estimating ${\rm v_{rot}}$, we assume, as~a first approximation, that the virial clouds are characterized by $<\tau>\simeq 1$, which means that the virial clouds are assumed to emit as black bodies at the CMB temperature. However, one needs to estimate the average optical depth of the virial clouds, $<\tau>$, over~a detector frequency range, $\nu_1-\nu_2$, which is, therefore, given as  $<\tau>=\displaystyle{\frac{1}{\nu_1-\nu_2}\int_{\nu_1}^{\nu_2}\tau_\nu~d\nu}$. The evaluation of this relation allows for determining whether the considered virial clouds are optically thick or thin. An alternative way to check the stability of the virial clouds is by estimating the ``stability time'' of these clouds, which is related to the probability of interaction of the molecules contaminating the virial clouds with the CMB photons. In \cite{qadir2019virial}, it was estimated the stability time for the virial clouds composed of purely ${\rm H_2}$ molecules. The estimated stability time results are much less than the collapsing time of the cloud, which showed that these clouds become stable before they could collapse. The considered situation was very simple, but when dust and heavier molecules are assumed to contaminate the clouds, the stability time and the collapsing time could vary because there could be quantum effects coming in, i.e., additional translational, rotational, and vibrational modes due to the interaction of the incoming radiation with heavier and complex molecules. 
\section{Halo~Model \label{halomodel}}
The aim of this section is to implement the generalized model we give in the previous sections, to~the considered galaxy halos. As~a first approximation, we assume that all the dark matter in the considered galactic halos is in the form of virial clouds. We also assume, for~definiteness, that a dust grain mass fraction contamination of the order of  1--3\% is present in the considered galaxy~halos.

We start by modeling the M31 galactic halo, and then we extend our analysis considering the other galaxies. Following \cite{tamm2012stellar}, we assume that the M31 halo size is $R \simeq 200$ kpc. However, the~reader should keep in mind that there is not a definite estimate for the size of the M31 halo. For~example, \cite{Lehner_2020} used ultraviolet absorption measurements of Si II, Si III, Si IV, C II, and~C IV from the Hubble Space Telescope/Cosmic Origins Spectrograph and O IV from the far Ultraviolet Spectrographic Explorer to estimate how metals are distributed in the circumgalactic medium (CGM) of M31 and found that the halo of M31 might extend up to $\approx$300 kpc.

\begin{figure}
        \centering
        \begin{subfigure}[t]{0.7\textwidth}
              \centering
            \includegraphics[scale=.7]{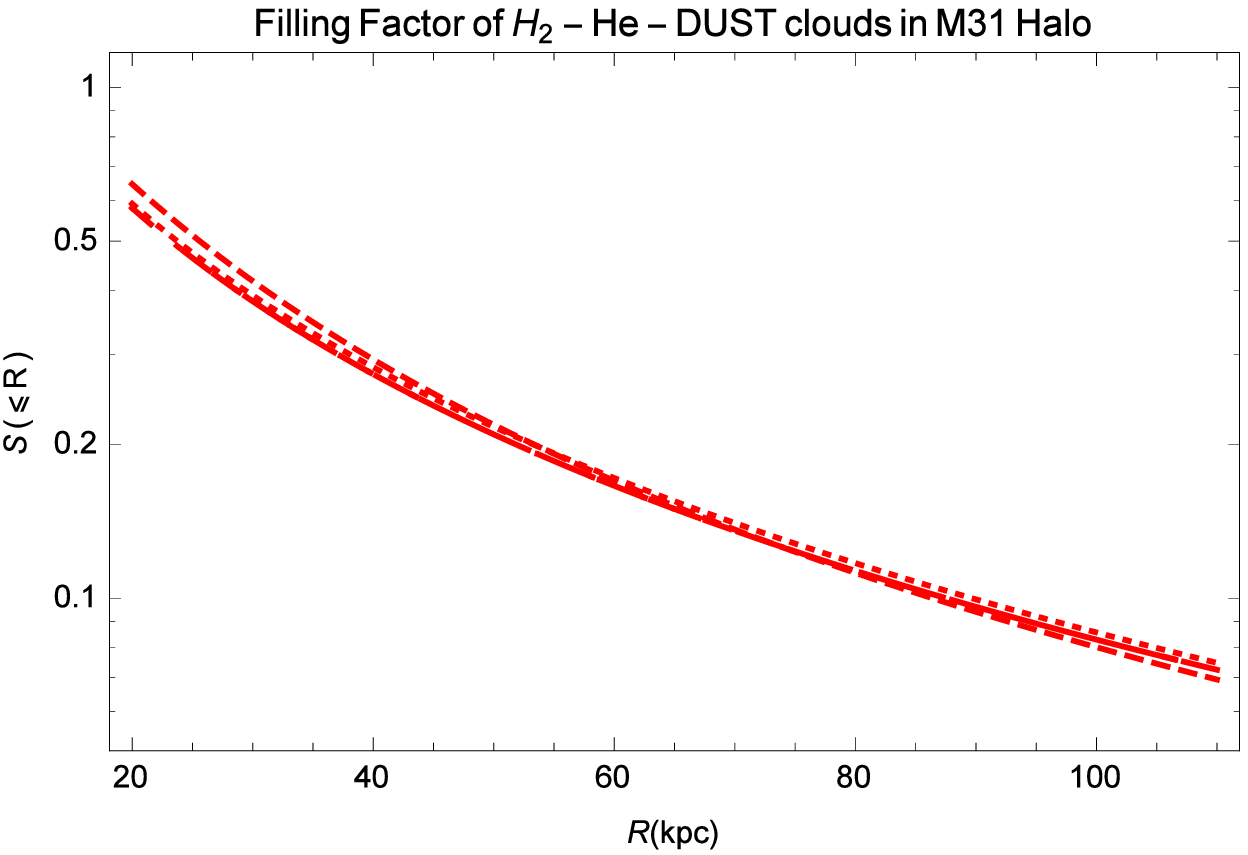}
            \caption{\label{figd}}
        \end{subfigure}
        \\
        \begin{subfigure}[t]{0.7\textwidth}
              \centering
            \includegraphics[scale=1]{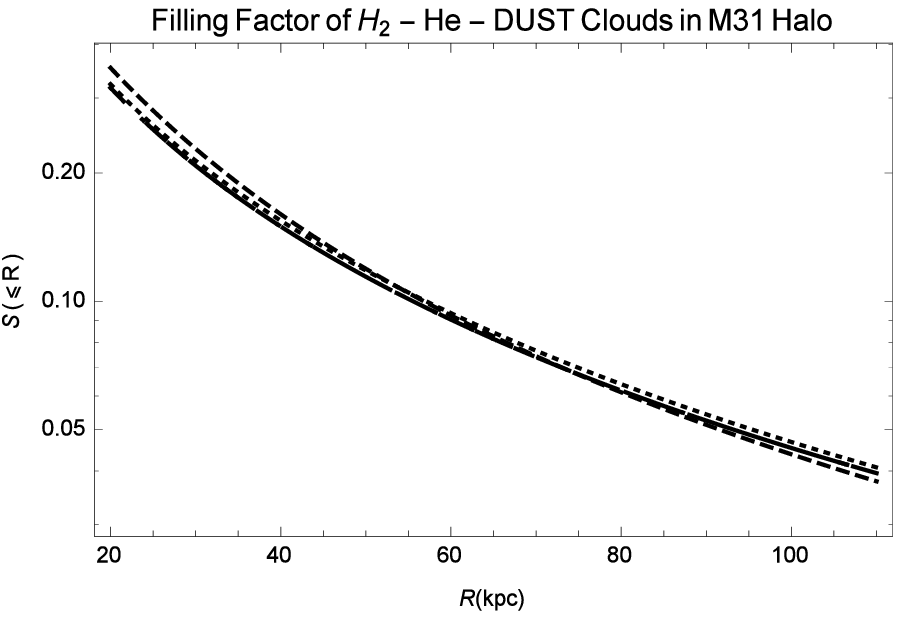}
            \caption{\label{fige}}
        \end{subfigure}
        ~
        \begin{subfigure}[t]{0.7\textwidth}
              \centering
            \includegraphics[scale=1]{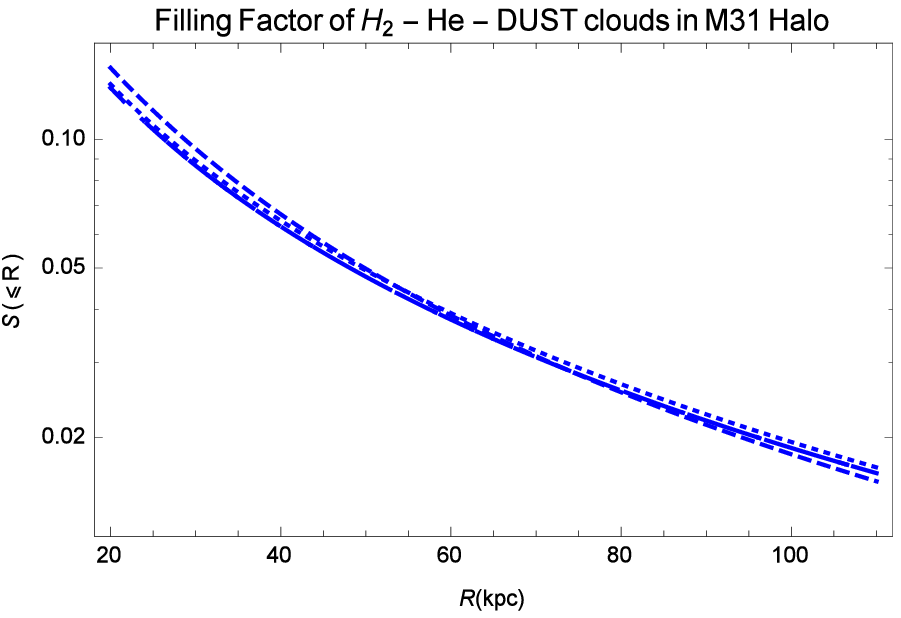}
            \subcaption{\label{figf}}
        \end{subfigure}
        \caption{The filling factor for the three fluids virial cloud model with 1\% (\textbf{a}), 2\% (\textbf{b}), and~3\% (\textbf{c}) dust contamination is given in the case of the M31 halo. The~bold, dotted, and dashed curves correspond to the NFW, Moore, and Burkert models, respectively. \label{figm31filling}}

    \end{figure}

In Figure~\ref{figm31filling}, we give the cloud filling factor obtained through Equation~(\ref{eq5}). It is seen that for all the considered virial cloud models, the filling factor turns out to be less than 1. It is also seen that the obtained values of the filling factor for the three dark matter models when we consider three fluids virial clouds are more or less the~same.

\begin{figure}
        \centering
        \begin{subfigure}[t]{0.7\textwidth}
              \centering
            \includegraphics[scale=.7]{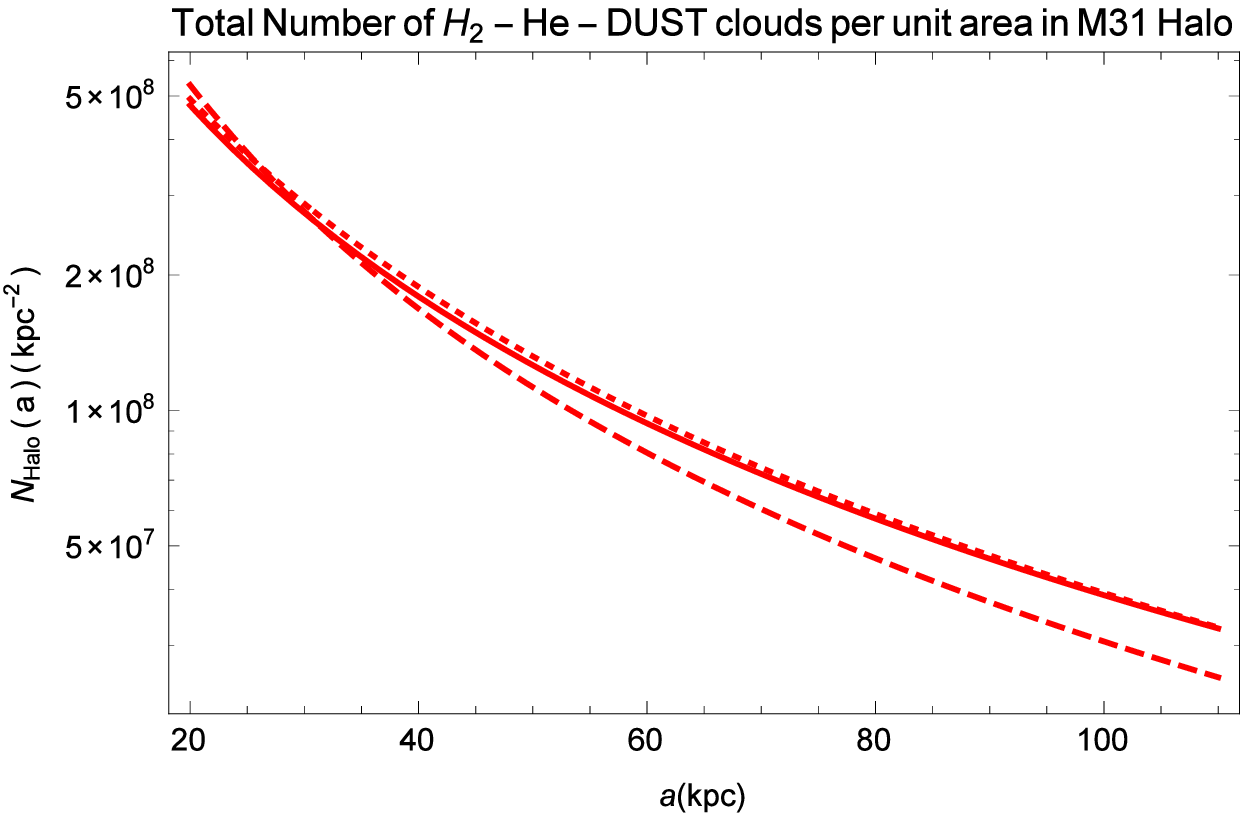}
            \caption{\label{figa}}
        \end{subfigure}%
        \\
        \begin{subfigure}[t]{0.7\textwidth}
               \centering
            \includegraphics[scale=.7]{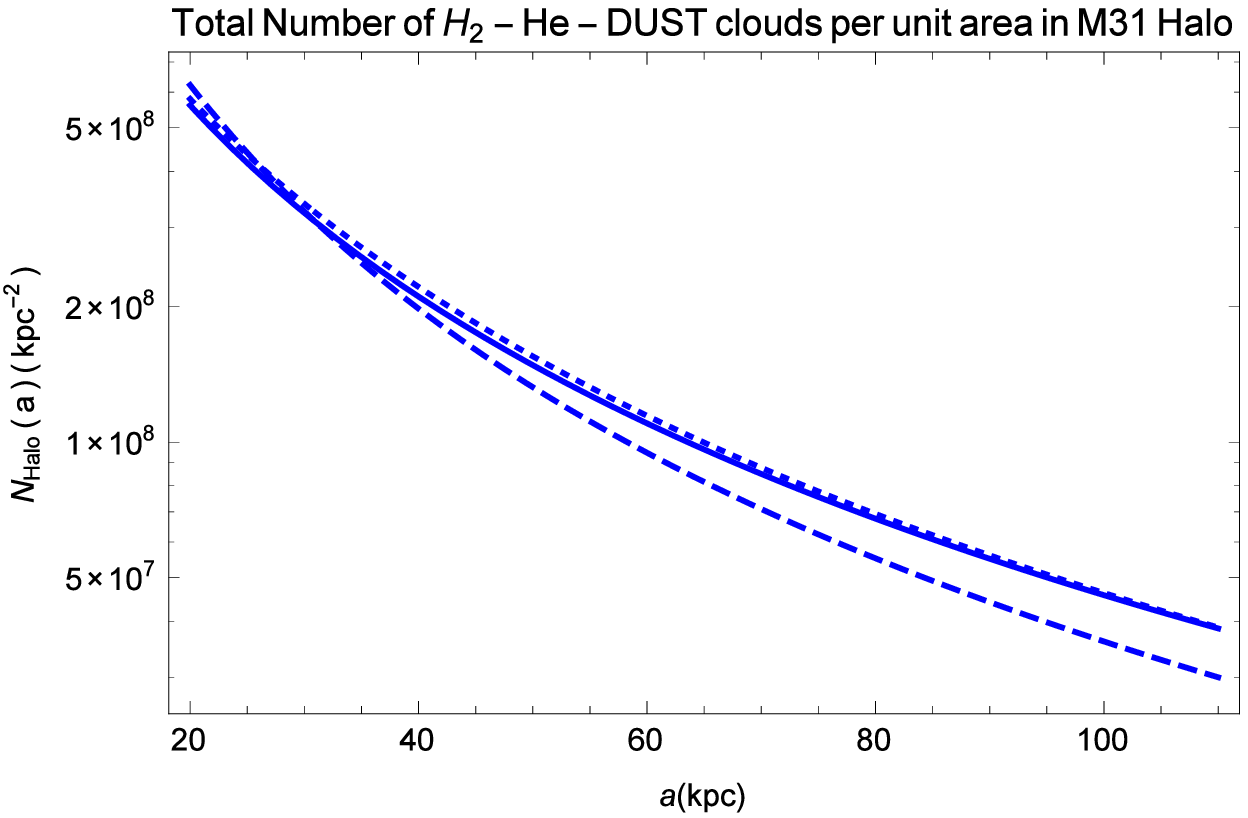}
            \caption{\label{figb}}
        \end{subfigure}
        ~
        \begin{subfigure}[t]{0.7\textwidth}
              \centering
            \includegraphics[scale=.7]{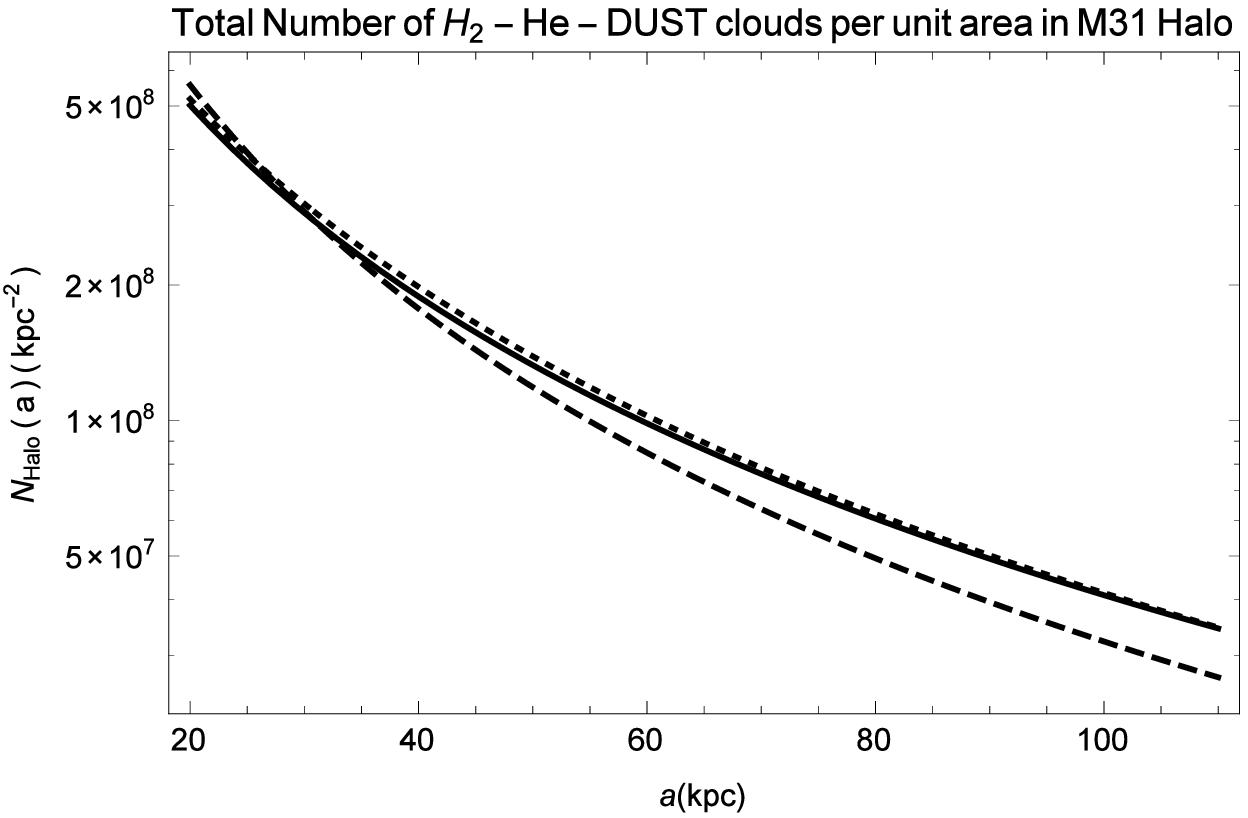}
            \caption{\label{figc}}
        \end{subfigure}
        \caption{The cumulative radial distribution  of the number of clouds per unit area for the three fluids virial cloud models with 1\% (\textbf{a}), 2\% (\textbf{b}), and~3\% (\textbf{c}) dust contamination is given in the case of the M31 halo. The~bold curve is for the NFW model, the~dotted curve corresponds to the Moore model, while the dashed one is for the Burkert~model. \label{figm31number}}

    \end{figure}

The virial cloud cumulative radial distribution towards the M31 halo, for~a dust concentration fraction in the range 1--3\%, is shown in Figure~\ref{figm31number}, assuming $f=1$. From~these plots, it is clear that the number of virial clouds in the M31 halo does not depend strongly on the considered halo model, but~it depends more heavily on the cloud chemical composition and dust~abundance.

We are now in a position to estimate the rotational velocity of the M31 halo, within~various galactocentric distances using Equation~(\ref{eq6}). We use the SMICA processed data to obtain the values of $\Delta T$ for M31 galaxy, which is already analyzed in \citep{tahir2019constraining}. The~SMICA data were chosen because they are less contaminated than the other available {\it Planck} bands. We estimated the rotational velocity values within 41.5 kpc to 103.8 kpc, and~the obtained results are shown in Table~\ref{tab3}.

We remark here that HI observations towards the Andromeda galaxy obtained with the Effelsberg and Green Bank 100 m telescopes showed that the rotation curve in the M31 disk up to a galactocentric distance $\approx 35$ kpc, is approximately flat with velocity $\approx 226$~km/s. Moreover, M31 rotation curves up to a  galactocentric distance $\approx 100$ kpc were built by combining disk rotation velocities and radial velocities of satellite galaxies and globular clusters. Ref.~\cite{carignan2006extended} adopted the NFW model to estimate these curves obtaining an approximately flat behavior with $v_{rot}\approx$  110--200 km/s. However, as~already noted, the~M31 halo rotation velocity is much less constrained by observations with respect to the disk rotational velocity.  In~particular, from~Table~\ref{tab3}, we see that for $f=1$, the~rotational velocity of the M31 halo follows a trend similar to that obtained for the disk, for~the virial cloud model with 1\% , 2\%, and~3\% contamination of dust grains. The~estimated M31 halo rotational velocity turns out to be ${\rm v_{rot}} \approx$ 50--230 km/s, at~large galactocentric distances (that is, in the region between about 20 kpc  and 108 kpc from the M31 center).
\begin{table}
\centering
\begin{tabular}{ |c|c|c|c|c|c|c|c|c|c|c|c| } 
\hline
\boldmath{$f_{dust}$} & \textbf{R} & \boldmath{$\Delta T/T$} &  & \boldmath{$N$} & &  & \boldmath{$S$} &  & & \boldmath{${\rm v_{rot}}$} &   \\
	&	\textbf{(kpc)} & \boldmath{$(10^{-6})$} &  & \boldmath{$10^8 ({\rm kpc^{-2}})$} & &  & \boldmath{$(10^{-2})$} &  & & \textbf{(km/s)} &   \\
		
	&	 &  & \boldmath{$N^N$} & \boldmath{$N^M$} & \boldmath{$N^B$} & \boldmath{$S_N$} & \boldmath{$S_M$} & \boldmath{$S_B$}  & \boldmath{${\rm v_{rot}^N}$} & \boldmath{${\rm v_{rot}^M}$} & \boldmath{${\rm v_{rot}^B}$}   \\
\hline
1\%&21.4 & $-$0.98 & 4.36 & 4.52  & 16.25 & 27.20 & 28.28 & 100 & $-$0.55& $-$0.53 & $-$0.99\\
&		31.1 & 2.36 & 2.60  & 2.73 & 5.67 & 16.38 & 17.10 & 35.4 & 2.23 & 2.12 & 3.27\\
&		41.8 & 6.40 & 1.69 & 1.77 & 7.90 & 10.57 & 11.11 & 49.41 & 9.41 & 8.96 & 12.28\\
&		51.8 & 14.3 &  1.18 & 1.24 & 6.16 & 7.42 & 7.75 & 38.50 & 29.86 & 28.57 & 28.27\\
&		77.8 & 7.90 & 0.67  & 0.61  & 3.93 &3.76  & 3.86 & 24.59 & 32.55 & 31.71 & 34.09 \\
&		103.8 & 7.80 & 0.36 & 0.36 & 2.87 & 2.22 & 2.29 & 17.95 & 53.34 & 52.87 & 37.28\\
2\%&		21.4 & $-$0.98 & 4.59 & 4.76 & 17.11 & 14.89 & 15.43 & 55.45 & $-$1.01 & $-$0.98 & $-$1.82\\
	&	31.1 & 2.36 & 2.74  & 2.88 & 5.97 & 8.99 & 9.33 & 19.34    & 4.08  & 3.89  & 6.00 \\
	&	41.8 & 6.40 & 1.78  & 1.87 & 8.32 & 5.77 & 6.05 & 26.96    & 17.25 & 16.43 & 22.59\\
	&	51.8 & 14.3 & 1.24  & 1.30 & 6.48 & 4.04 & 4.23 & 21.01    & 54.72 & 52.35 & 51.82\\
	&	77.8 & 7.90 & 0.63  & 0.65 & 4.14 & 2.20 & 2.11 & 13.42    & 59.65 & 58.12 & 62.48 \\
	&	103.8 & 7.80 & 0.38 & 0.31 & 3.02 & 1.24 & 1.25 & 9.79     & 97.76 & 96.90 & 68.32\\
3\% &			21.4 & $-$0.98 & 5.13 & 5.32 & 19.12 & 6.21 & 6.44 & 23.14 & $-$2.43  & $-$2.35  & $-$4.36\\
	&	31.1 & 2.36  & 3.06 & 3.21 & 6.67  & 3.71 & 3.89 & 8.08  & 9.79   & 9.33   & 14.39\\
	&	41.8 & 6.40  & 1.99 & 2.90 & 9.30  & 2.40 & 2.52 & 11.25 & 41.34  & 39.37  & 53.92\\
	&	51.8 & 14.3  & 1.39 & 1.45 & 7.24  & 1.69 & 1.76 & 8.77  & 131.10 & 125.42 & 119.69\\
	&	77.8 & 7.90  & 0.70 & 0.72  & 4.62 & 0.88 & 0.82 & 5.60  & 142.93 & 139.24 & 124.15 \\
	&	103.8 & 7.80 & 0.42 & 0.43 & 3.37  & 0.65 & 0.52 & 4.08  & 234.22 & 232.66 & 163.69\\
\hline
\end{tabular}
\caption{Estimated physical parameters of the M31 halo, populated by virial clouds with 1\%, 2\%, and~3\% dust~contamination. Note: The temperature asymmetry divided by the CMB temperature (column 3), the~total number of virial clouds per unit area within the observed radius (columns 4--6), the~filling factor (columns 7--9) and the estimated rotational velocity of the M31 halo (columns 10--12) are given, with~respect to the considered halo radii (column 2), for~the three considered dark matter models. \label{tab3}}
\end{table}

Similarly, in~the case of other spiral galaxies, we can apply the same procedure like M31 for estimating the halo rotational velocity of the other considered galaxies, namely M33, M81, M82, NGC 5128, and~NGC 4594, respectively. We also assumed that 1--3\% of dust contaminate virial clouds present in the halos of these galaxies{, and~$f=1$ is used as a reference value in Tables~\ref{tab3}--\ref{tab7}}.

The total number of virial clouds per unit area, the~filling factor of the clouds in each considered galaxy halo, and~the halo rotational velocity are shown in Tables~\ref{tab3}--\ref{tab7}. We have used the temperature asymmetry data published in the literature (see in particular the \cite{de2014planck, de2015planck, gurzadyan2015planck, de2016triangulum, gurzadyan2018messier, de2019rotating}) in the 70, 100, and~143 GHz bands in the case of M33, M81, M82, and~NGC 5128, and~SMICA band for NGC 5494 at a resolution corresponding to $N_{side}= 2048$, in~HEALPix scheme~\cite{gorski2005healpix}. The~temperature excesses are shown in Tables~\ref{tab8} and \ref{tab4} (see column 3 for NGC 5494 and column 4 for M33, M81, M82, and~NGC 5128, respectively). {Tables~\ref{tab3}--\ref{tab7} report the obtained values of the halo rotational velocity ${\rm v_{rot}}$  for the various adopted models and for $f_{dust}$ ranging from 1\% to 3\%. As~one can see, the~obtained ${\rm v_{rot}}$ values are in substantial agreement with the expectations. However, in~the case of some galaxies, there are regions in which apparently inconsistent values of ${\rm v_{rot}}$ are present. Considering in detail each analyzed galaxy, in~the case of the M31 halo (see Table~\ref{tab3}), the~rotational velocity is acceptable for all three adopted models and for all dust fractions. In~the case of NGC 5494, M33, and~M81 halo, respectively (see Tables~\ref{tab8}--\ref{tab5}), the~${\rm v_{rot}}$ values appear acceptable for $f_{dust}$ of 1\% and 2\% and only marginally acceptable for $f_{dust}=3\%$, as~a result, we believe that it is not likely to say that $f_{dust}<3\%$ for these galaxies. In~the case of the M82 halo (see Table~\ref{tab6}) at a large radius for all the dust concentrations, the ${\rm v_{rot}}$ values are not acceptable hence we can say that at larger radius $f_{dust}<1\%$. In~the case of NGC 5128 halo (see Table~\ref{tab7}), the rotational velocity values appear acceptable for all the~cases. 

Notice that the obtained ${\rm v_{rot}}$ values depend on the following assumptions: (1) We assume a constant dust fraction in each galaxy, independent of the galactocentric distance. However, as~is explained previously that the dust grain temperature can easily be higher. It should be borne in mind that dust grains in the halo follow a distribution profile that depends upon the galactocentric distance, and~to treat this model more physically, one needs to see how these dust grains are distributed in the halos; (2) the temperature of the halo dust grains is assumed to be exactly $2.7$ K (see \cite{yershov2020distant}), and~hence one needs to see how the temperature profile of dust changes with the galactocentric distance; (3) the distribution profile of the clouds in the halos of each considered galaxy: it can be seen that the NFW model gives more consistent results than the other two models, and (5) our assumption that the observed asymmetry in the CMB, i.e.,~the values of $\Delta T/T$ are solely due to virial clouds in the halo of each galaxy. However, there are other effects from a theoretical point of view that give can give rise to this asymmetry in the CMB. To treat the problem of the halo rotation more physically, one has to see how the halo dust is distributed and the chemical composition of the virial clouds, for each of the above galaxies separately, and~consider other possible contributing effects.

\begin{table}
\centering
\begin{tabular}{ |c|c|c|c|c|c|c|c|c|c|c|c| } 
\hline
\boldmath{$f_{dust}$} & \textbf{R} & \boldmath{$\Delta T/T$} &  &\textbf{ $N$} & &  & \boldmath{$S$} &  & & \boldmath{${\rm v_{rot}}$} &   \\
     	&	\textbf{(kpc)} & \boldmath{$(10^{-6})$} &  & \boldmath{$10^8 ({\rm kpc^{-2}})$} & &  & \boldmath{$(10^{-2})$} &  & & \textbf{(km/s)} &   \\
		
 & &  & \boldmath{$N^N$} & \boldmath{$N^M$} & \boldmath{$N^B$} & \boldmath{$S_N$} & \boldmath{$S_M$} & \boldmath{$S_B$}  & \boldmath{${\rm v_{rot}^N}$} & \boldmath{${\rm v_{rot}^M}$} & \boldmath{${\rm v_{rot}^B}$}   \\
\hline
 1\% & 33.4 & $-$8.44 & 20.26 & 15.73 & 12.43 & 12.66 & 19.86 & 17.74 & $-$10.37 & $-$12.92 & $-$17.25\\
       &		66.8 & $-$6.60 & 5.58  & 4.13 & 6.21 & 3.49 & 2.58 & 4.43 & $-$28.53 & $-$38.65 & $-$42.44 \\
       &		100.2 & 0.36 & 2.56 & 1.85 & 4.87 & 1.60 & 1.16 & 3.13 & 34.54 & 47.75 & 49.14\\
       &		167 & 3.66 & 0.95 & 0.67 & 0.57 & 0.59 & 0.42 & 0.18 & 39.32 & 33.67 & 49.58\\
       2\%& 33.4 & $-$8.44 & 21.32 & 16.62 & 13.82 & 6.91 & 5.93 & 4.83 & $-$18.45 & $-$23.68 & $-$23.71\\
       &		66.8 & $-$6.60 & 5.88  & 4.34 & 6.97 & 1.90 & 1.40 & 1.22 & $-$52.33 & $-$70.84 & $-$60.91 \\
       &		100.2 & 0.36 & 2.70 & 1.95 & 5.13 & 0.87 & 0.63 & 0.64 & 63.30 & 87.51 & 92.91\\
       &		167 & 3.66 & 1.00 & 0.70 & 0.30 & 0.32 & 0.22 & 0.09 & 71.00 & 84.25 & 88.93\\
       3\% & 33.4 & $-$8.44 & 23.83 & 18.57 & 14.62 & 2.88 & 2.22 & 1.72 & $-$44.21 & $-$56.74 & $-$31.42\\
       &		66.8 & $-$6.60 & 6.57  & 4.86 & 7.79 & 0.79 & 0.58 & 0.92 & $-$125.50 & $-$169.72 & $-$143.44 \\
       &		100.2 & 0.36 & 3.02 & 2.18 & 5.74 & 0.36 & 0.26 & 0.68 & 151.65 & 200.96 & 160.91\\
       &		167 & 3.66 & 1.11 & 0.79 & 0.32 & 0.13 & 0.09 & 0.04 & 140.74 & 158.01 & 164.51\\
\hline
\end{tabular}
\caption{Same as Table~\ref{tab3}, but~for NGC 5494~halo. Note: All columns have the same meaning as Table~\ref{tab3}, but~for NGC 5494 halo. \label{tab8}}
\end{table}

\begin{landscape}
\begin{table}
\centering
\begin{tabular}{ |c|c|c|c|c|c|c|c|c|c|c|c|c| } 
\hline
\boldmath{$\nu$} & \boldmath{$f_{dust}$} & \textbf{R} & \boldmath{$\Delta T/T$} &  & \boldmath{$N$} & &  & \boldmath{$S$} &  & &\textbf{ ${\rm v_{rot}}$} &   \\
   \textbf{(GHz)}  &   &	\textbf{(kpc)} & \boldmath{$(10^{-6})$} &  & \boldmath{$10^8 ({\rm kpc^{-2}})$} & &  & \boldmath{$(10^{-2})$} &  & & \textbf{(km/s)} &   \\
		
 &	 &  & & \boldmath{$N^N$} & \boldmath{$N^M$} & \boldmath{$N^B$} & \boldmath{$S_N$} & \boldmath{$S_M$} & \boldmath{$S_B$}  & \boldmath{${\rm v_{rot}^N}$} &\textbf{ ${\rm v_{rot}^M}$} & \boldmath{${\rm v_{rot}^B}$}   \\
\hline
 70  & 1\%&92 & 25.68 & 3.13 & 7.26 & 1.35 & 19.20 & 4.50 & 8.46 & 22.92 & 98.95 & 53.17\\
     & &		171 & 11.0 & 1.12  & 2.18 & 5.28 & 7.02 & 1.74 & 3.34 & 27.36 & 141.36 & 31.49\\
     & &		245 & 5.13 & 0.59 & 1.07 & 3.60 & 3.57 & 6.71 & 2.21 & 24.12 & 134.08 & 21.10\\
     & 2\%&		92 & 25.68 & 3.30 & 7.65 & 1.42 & 17.89 & 2.43 & 4.65 & 42.04 & 181.33 & 97.45\\
     &	&	171 & 11.0 & 1.18  & 2.29 & 5.55 & 3.99 & 11.33 & 8.14 & 50.14  & 259.06  & 57.81 \\
     & 	&	245 & 5.13 & 6.27  & 1.12 & 3.79 & 20.77 & 3.05 & 1.96    & 44.20 & 245.71 & 38.67\\
     & 3\% &	92 & 21.4 & 3.69 & 8.55 & 1.59 & 4.21 & 0.13 & 0.19 & 100.63  & 434.35  & 233.47\\
     & &	171 & 11.0  & 1.32 & 2.56 & 6.21 & 1.71 & 0.38 & 0.075  & 120.14   & 620.68   & 138.51\\
     &	&	245 & 5.13  & 7.01 & 1.26 & 4.27  & 0.89 & 0.015 & 0.0052 & 105.91  & 588.67  & 92.64\\
 100 & 1\%&92 & 22.01 & 3.13 & 7.26 & 1.35 & 19.20 & 4.50 & 8.46 & 19.64 & 84.81 & 45.58\\
     & &		171 & 15.41 & 1.12  & 2.18 & 5.28 & 7.02 & 1.74 & 3.34 & 38.30 & 197.91 & 44.16\\
     & &		245 & 6.60& 0.59 & 1.07 & 3.60 & 3.57 & 6.71 & 2.21 & 31.01& 172.38 & 27.13\\
     & 2\%&		92 & 22.01 & 3.30 & 7.65 & 1.42 & 17.89 & 2.43 & 4.65 & 36.01 & 135.42 & 83.52\\
     &	&	171 & 15.41 & 1.18  & 2.29 & 5.55 & 3.99 & 11.33 & 8.14 & 70.20  & 362.69  & 80.94 \\
     & 	&	245 & 6.60 & 6.27  & 1.12 & 3.79 & 20.77 & 3.05 & 1.96    & 56.83 & 315.91 & 49.71\\
     & 3\% &	92 & 22.01 & 3.69 & 8.55 & 1.59 & 4.21 & 0.13 & 0.19 & 86.25  & 372.37  & 200.17\\
     & &	171 & 15.41  & 1.32 & 2.56 & 6.21 & 1.71 & 0.38 & 0.075  & 168.19   & 868.95   & 193.92\\
     &	&	245 & 6.60  & 7.01 & 1.26 & 4.27  & 0.89 & 0.015 & 0.0052 & 136.17  & 756.87  & 119.11\\
143  & 1\%&92 & 21.46 & 3.13 & 7.26 & 1.35 & 19.20 & 4.50 & 8.46 & 19.15 & 82.69 & 44.44\\
     & &		171 & 16.87 & 1.12  & 2.18 & 5.28 & 7.02 & 1.74 & 3.34 & 41.95 & 216.76 & 48.37\\
     & &		245 & 7.15 & 0.59 & 1.07 & 3.60 & 3.57 & 6.71 & 2.21 & 33.60 & 186.75 & 29.39\\
     & 2\%&		92 & 21.46& 3.30 & 7.65 & 1.42 & 17.89 & 2.43 & 4.65 & 35.10 & 151.54 & 81.44\\
     &	&	171 & 16.87& 1.18  & 2.29 & 5.55 & 3.99 & 11.33 & 8.14 & 76.89 & 397.23  & 88.65 \\
     & 	&	245 & 7.15& 6.27  & 1.12 & 3.79 & 20.77 & 3.05 & 1.96    & 61.57 & 342.23 & 53.86\\
     & 3\% &	92 & 21.46 & 3.69 & 8.55 & 1.59 & 4.21 & 0.13 & 0.19 & 84.10  & 363.06  & 195.11\\
     & &	171 & 16.87  & 1.32 & 2.56 & 6.21 & 1.71 & 0.38 & 0.075  & 184.21   & 951.71   & 212.39\\
     &	&	245 & 7.15  & 7.01 & 1.26 & 4.27  & 0.89 & 0.015 & 0.0052 & 147.52  & 819.94  & 129.04\\
\hline
\end{tabular}
\caption{Estimated physical parameters of the clouds in the M33 halo and its rotational~velocity. Note: We give, for~the considered Planck frequency band (column 1), and~for each considered  halo radii (column 3), the~temperature asymmetry divided by the CMB temperature (column 4), the~total number of virial clouds per unit area within the observed radius (column 5--7), the~filling factor (column 8--10) and the estimated rotational velocity of the M31 halo (column 11--13), for~the three considered dark matter models. \label{tab4}}
\end{table}
\end{landscape}

\begin{landscape}
\begin{table}
\centering
\begin{tabular}{ |c|c|c|c|c|c|c|c|c|c|c|c|c| } 
\hline
\boldmath{$\nu$} & \boldmath{$f_{dust}$} & \textbf{R} & \boldmath{$\Delta T/T$} &  & \boldmath{$N$} & &  & \boldmath{$S$} &  & &\textbf{ ${\rm v_{rot}}$} &   \\
   \textbf{(GHz)}  &   &	\textbf{(kpc)} & \boldmath{$(10^{-6})$} &  & \boldmath{$10^8 ({\rm kpc^{-2}})$} & &  & \boldmath{$(10^{-2})$} &  & & \textbf{(km/s)} &   \\
		
 &	 &  & & \boldmath{$N^N$} & \boldmath{$N^M$} & \boldmath{$N^B$} & \boldmath{$S_N$} & \boldmath{$S_M$} & \boldmath{$S_B$}  & \boldmath{${\rm v_{rot}^N}$} &\textbf{ ${\rm v_{rot}^M}$} & \boldmath{${\rm v_{rot}^B}$}   \\
\hline
 70  & 1\% & 15 & 20.36 & 14.80 & 15.56 & 18.85 & 0.92 & 0.97 & 0.98 & 13.12 & 12.90 & 48.44\\
     & &		30 & 23.85 & 5.28  & 5.49 & 13.39 & 0.33 & 0.34 & 0.87 & 44.51 & 43.08 & 106.45\\
     & &		60 & 29.35 & 1.63 & 1.62 & 4.60 & 0.10 & 0.10 & 0.28 & 17.84 & 17.91 & 25.8 \\
     & 2\%& 15 & 20.36 & 15.60 & 16.37 & 13.37 & 0.50 & 0.53 & 0.98 & 24.44 & 23.79 & 88.76\\
     & &		30 & 23.85 & 5.56  & 5.78 & 14.47 & 0.18 & 0.18 & 0.47 & 82.09 & 78.94 & 195.08\\
     & &		60 & 29.35 & 1.71 & 1.71 & 4.84 & 0.055 & 0.055 & 0.15 & 32.69 & 32.83 & 47.30 \\
     & 3\% & 15 & 20.36 & 17.40 & 18.30 & 13.39 & 0.21 & 0.22 & 0.41 & 54.60 & 57.01 & 212.67\\
     & &		30 & 23.85 & 6.21  & 6.46 & 16.64 & 0.072 & 0.078 & 0.019 & 194.64 & 189.14 & 467.38\\
     & &		60 & 29.35 & 1.92 & 1.91 & 5.41 & 0.023 & 0.023 & 0.063 & 78.33 & 78.67 & 113.97 \\
 100  & 1\%& 15 & 14.67 & 14.80 & 15.56 & 18.85 & 0.92 & 0.97 & 0.98 & 19.81 & 19.31 & 34.91\\
     & &		30 & 20.18 & 5.28  & 5.49 & 13.39 & 0.33 & 0.34 & 0.87 & 37.86 & 36.45 & 90.07 \\
     & &		60 & 27.52 & 1.63 & 1.62 & 4.60 & 0.10 & 0.10 & 0.28 & 36.26 & 36.79 & 24.23 \\
     & 2\%& 15 & 14.67 & 15.60 & 16.37 & 13.37 & 0.50 & 0.53 & 0.98 & 17.97 & 17.15 & 63.97\\
     & &		30 & 20.18 & 5.56  & 5.78 & 14.47 & 0.18 & 0.18 & 0.47 & 69.39 & 66.80 & 165.06\\
     & &		60 & 27.52 & 1.71 & 1.71 & 4.84 & 0.055 & 0.055 & 0.15 & 30.52 & 30.78 & 44.11 \\
     & 3\% & 15 & 14.67 & 17.40 & 18.30 & 13.39 & 0.21 & 0.22 & 0.41 & 43.07 & 41.08 & 153.28\\
     & &		30 & 20.18 & 6.21  & 6.46 & 16.64 & 0.072 & 0.078 & 0.019 & 166.25 & 160.04 & 395.47\\
     & &		60 & 27.52 & 1.92 & 1.91 & 5.41 & 0.023 & 0.023 & 0.063 & 73.43 & 73.75 & 106.40 \\
143   & 1\%& 15 & 16.51 & 14.80 & 15.56 & 18.85 & 0.92 & 0.97 & 0.98 & 11.03 & 10.52 & 39.27\\
     & &		30 & 18.34 & 5.28  & 5.49 & 13.39 & 0.33 & 0.34 & 0.87 & 34.42 & 33.13 & 81.88 \\
     & &		60 & 27.52 & 1.63 & 1.62 & 4.60 & 0.10 & 0.10 & 0.28 & 16.72 & 16.79 & 24.23 \\
     & 2\%& 15 & 16.51 & 15.60 & 16.37 & 13.37 & 0.50 & 0.53 & 0.98 & 20.22 & 19.29 & 71.97 \\
     & &		30 & 18.34 & 5.56  & 5.78 & 14.47 & 0.18 & 0.18 & 0.47 & 63.08 & 60.72 & 150.06\\
     & &		60 & 27.52 & 1.71 & 1.71 & 4.84 & 0.055 & 0.055 & 0.15 & 30.65 & 30.78 & 44.41 \\
     & 3\% & 15 & 16.51 & 17.40 & 18.30 & 13.39 & 0.21 & 0.22 & 0.41 & 48.45 & 46.22 & 172.44 \\
     & &		30 & 18.34 & 6.21  & 6.46 & 16.64 & 0.072 & 0.078 & 0.019 & 151.13 & 145.49 & 359.54 \\
     & &		60 & 27.52 & 1.92 & 1.91 & 5.41 & 0.023 & 0.023 & 0.063 & 73.43 & 73.75 & 106.40 \\
\hline
\end{tabular}
\caption{Same as Table~\ref{tab4} but for M81~halo. Note:  All columns have the same meaning as Table~\ref{tab4} but for M81 galaxy halo. \label{tab5}}
\end{table}
\end{landscape}

\begin{landscape}
\begin{table}
\centering
\begin{tabular}{ |c|c|c|c|c|c|c|c|c|c|c|c|c| } 
\hline
\boldmath{$\nu$} & \boldmath{$f_{dust}$} & \textbf{R} & \boldmath{$\Delta T/T$} &  & \boldmath{$N$} & &  & \boldmath{$S$} &  & &\textbf{ ${\rm v_{rot}}$} &   \\
   \textbf{(GHz)}  &   &	\textbf{(kpc)} & \boldmath{$(10^{-6})$} &  & \boldmath{$10^8 ({\rm kpc^{-2}})$} & &  & \boldmath{$(10^{-2})$} &  & & \textbf{(km/s)} &   \\
		
 &	 &  & & \boldmath{$N^N$} & \boldmath{$N^M$} & \boldmath{$N^B$} & \boldmath{$S_N$} & \boldmath{$S_M$} & \boldmath{$S_B$}  & \boldmath{${\rm v_{rot}^N}$} &\textbf{ ${\rm v_{rot}^M}$} & \boldmath{${\rm v_{rot}^B}$}   \\
\hline
  70  & 1\% & 15 & 18.16 & 14.57 & 16.61 & 16.74 & 9.10 & 10.03 & 10.46 & 68.25 & 59.85 & 118.15\\
     & &		30 & 15.59 & 5.31  & 5.35 & 13.02 & 3.32 & 3.34 & 8.13 & 160.69 & 159.58 & 213.09\\
     & &		60 & $-$5.87 & 1.66 & 1.49 & 8.03 & 1.04 & 0.93 & 5.04 & $-$192.72 & $-$214.66 & $-$163.03\\
     & 2\%&		15 & 18.16 & 15.33 & 17.49 & 17.63 & 4.96 & 5.66 & 5.71 & 125.07 & 109.68 & 216.53\\
     &	&	30 & 15.59 & 5.59  & 5.63 & 13.70 & 1.82 & 1.82 & 4.44 & 294.49  & 294.47  & 390.51 \\
     & 	&	60 & $-$5.87 & 1.75  & 1.57 & 8.46 & 0.56 & 0.51 & 2.74  & $-$353.18 & $-$393.38 & $-$298.78\\
     & 3\% &	15 & 18.16 & 17.14 & 19.54 & 19.70 & 2.07 & 2.36 & 2.38 & 299.66  & 262.78  & 518.76\\
     & &	30 & 15.59 & 6.25 & 6.29 & 15.32 & 0.76 & 0.76 & 1.85  & 705.54   & 700.65   & 935.60\\
     & & 60 & $-$5.87  & 1.96 & 1.76 & 9.45  & 0.23 & 0.21 & 1.14 & $-$846.17  & $-$942.46  & $-$715.82\\
100  & 1\%& 15 & 16.88 & 14.57 & 16.61 & 16.74 & 9.10 & 10.03 & 10.46 & 55.62 & 109.80 & 109.80 \\
     & &		30 & 17.61 & 5.31  & 5.35 & 13.02 & 3.32 & 3.34 & 8.13 & 181.49 & 180.23 & 240.67 \\
     & &		60 & $-$6.60 & 1.66 & 1.49 & 8.03 & 1.04 & 0.93 & 5.04 & $-$216.87 & $-$241.49 & $-$183.34\\
     & 2\%&		15 & 16.88 & 15.33 & 17.49 & 17.63 & 4.96 & 5.66 & 5.71 & 116.23 & 101.92 & 201.22 \\
     &	&	30 & 17.61 & 5.59  & 5.63 & 13.70 & 1.82 & 1.82 & 4.44 & 332.60  & 330.29  & 441.05 \\
     & 	&	60 & $-$6.60 & 1.75  & 1.57 & 8.46 & 0.56 & 0.51 & 2.74  & $-$397.35 & $-$442.52 & $-$336.12 \\
     & 3\% &	15 & 16.88 & 17.14 & 19.54 & 19.70 & 2.07 & 2.36 & 2.38 & 278.47  & 244.20  & 482.20 \\
     & &	30 & 17.61 & 6.25 & 6.29 & 15.32 & 0.76 & 0.76 & 1.85  & 796.85   & 791.32   & 566.68 \\
     & & 60 & $-$6.60  & 1.96 & 1.76 & 9.45  & 0.23 & 0.21 & 1.14 & $-$951.94  & $-$900.60  & $-$805.30\\
 143   & 1\%& 15 & 22.01 & 14.57 & 16.61 & 16.74 & 9.10 & 10.03 & 10.46 & 82.73 & 72.54 & 143.22 \\
     & &		30 & 22.01 & 5.31  & 5.35 & 13.02 & 3.32 & 3.34 & 8.13 & 226.86 & 225.29 & 300.84 \\
     & &		60 & $-$4.40 & 1.66 & 1.49 & 8.03 & 1.04 & 0.93 & 5.04 & $-$144.54 & $-$160.99 & $-$122.28 \\
     & 2\%&		15 & 22.01 & 15.33 & 17.49 & 17.63 & 4.96 & 5.66 & 5.71 & 151.60 & 132.95 & 262.46\\
     &	&	30 & 22.01 & 5.59  & 5.63 & 13.70 & 1.82 & 1.82 & 4.44 & 415.75  & 412.86  & 551.31 \\
     & 	&	60 & $-$4.40 & 1.75  & 1.57 & 8.46 & 0.56 & 0.51 & 2.74  & $-$264.89 & $-$295.03 & $-$224.08\\
     & 3\% &	15 & 22.01 & 17.14 & 19.54 & 19.70 & 2.07 & 2.36 & 2.38 & 363.22  & 318.52  & 268.81 \\
     & &	30 & 22.01 & 6.25 & 6.29 & 15.32 & 0.76 & 0.76 & 1.85  & 996.06   & 989.15   & 932.08 \\
     & & 60 & $-$4.40  & 1.96 & 1.76 & 9.45  & 0.23 & 0.21 & 1.14 & $-$634.63  & $-$706.85  & $-$736.68\\
\hline
\end{tabular}
\caption{Same as Table~\ref{tab4} but for the case of M82~halo. Note: All columns have the same meaning as Table~\ref{tab4} but for M82 galaxy halo. \label{tab6}}
\end{table}
\end{landscape}

\begin{landscape}
\begin{table}
\centering
\begin{tabular}{ |c|c|c|c|c|c|c|c|c|c|c|c|c| } 
\hline
\boldmath{$\nu$} & \boldmath{$f_{dust}$} & \textbf{R} & \boldmath{$\Delta T/T$} &  & \boldmath{$N$} & &  & \boldmath{$S$} &  & &\textbf{ ${\rm v_{rot}}$} &   \\
   \textbf{(GHz)}  &   &	\textbf{(kpc)} & \boldmath{$(10^{-6})$} &  & \boldmath{$10^8 ({\rm kpc^{-2}})$} & &  & \boldmath{$(10^{-2})$} &  & & \textbf{(km/s)} &   \\
		
 &	 &  & & \boldmath{$N^N$} & \boldmath{$N^M$} & \boldmath{$N^B$} & \boldmath{$S_N$} & \boldmath{$S_M$} & \boldmath{$S_B$}  & \boldmath{${\rm v_{rot}^N}$} &\textbf{ ${\rm v_{rot}^M}$} & \boldmath{${\rm v_{rot}^B}$}   \\
\hline
70  & 1\% & 92 & 25.68 & 18.68 & 12.75 & 13.48 & 11.16 & 7.97 & 8.42 & 13.09 & 19.16 & 11.22\\
    & &		171 & 11.74 & 6.22  & 3.76 & 6.02 & 3.89 & 2.35 & 3.76 & 17.94 & 29.67 & 38.63 \\
    & &		245 & 4.40 & 3.19 & 1.84 & 4.74 & 1.99 & 1.15 & 2.96 & 13.14 & 22.72 & 46.30\\
    & 2\%&		92 & 25.68 & 19.66 & 13.43 & 14.19 & 6.37 & 4.35 & 4.59 & 23.99 & 35.13 & 18.57\\
    &	&	171 & 11.74 & 6.55  & 3.96 & 6.33 & 2.12 & 1.28 & 2.05 & 32.89  & 54.38  & 57.84 \\
    & 	&	245 & 4.40 & 3.36  & 1.94 & 4.99 & 1.09 & 0.62 & 1.62  & 24.01 & 41.63 & 84.85 \\
    & 3\% &	92 & 25.68 & 21.97 & 15.01 & 15.86 & 2.67 & 1.83 & 1.91 & 57.48  & 84.16  & 44.49 \\
    & &	171 & 11.74  & 7.32 & 4.43 & 7.08 & 0.88 & 0.53 & 0.85  & 78.80   & 130.30   & 78.81 \\
    &	&	245 & 4.40 & 3.76 & 2.17 & 5.58  & 0.45 & 0.26 & 0.67 & 57.53  & 99.75  & 120.33 \\
100   & 1\%& 92 & 22.01 & 18.68 & 12.75 & 13.48 & 11.16 & 7.97 & 8.42 & 11.22 & 16.43 & 18.68\\
     & &		171 & 15.41 & 6.22  & 3.76 & 6.02 & 3.89 & 2.35 & 3.76 & 23.55 & 38.95 & 30.46 \\
     & &		245 & 6.60 & 3.19 & 1.84 & 4.74 & 1.99 & 1.15 & 2.96 & 19.65 & 34.08 & 69.45\\
     & 2\%&		92 & 22.01 & 19.66 & 13.43 & 14.19 & 6.37 & 4.35 & 4.59 & 20.56 & 30.11 & 15.94\\
     &	&	171 & 15.41 & 6.55  & 3.96 & 6.33 & 2.12 & 1.28 & 2.05 & 71.38  & 120.71  & 47.28 \\
     & 	&	245 & 6.60 & 3.36  & 1.94 & 4.99 & 1.09 & 0.62 & 1.62  & 62.45 & 127.85 & 139.02 \\
     & 3\% &	92 & 22.01 & 21.97 & 15.01 & 15.86 & 2.67 & 1.83 & 1.91 & 72.14  & 68.13  & 77.63 \\
     & &	171 & 15.41  & 7.32 & 4.43 & 7.08 & 0.88 & 0.53 & 0.85  & 171.02   & 149.63   & 113.27 \\
     &	&	245 & 6.60 & 3.76 & 2.17 & 5.58  & 0.45 & 0.26 & 0.67 & 130.49  & 93.49  & 97.36 \\
143   & 1\%& 92 & 21.28 & 18.68 & 12.75 & 13.48 & 11.16 & 7.97 & 8.42 & 10.84 & 15.83 & 18.39\\
     & &		171 & 16.88 & 6.22  & 3.76 & 6.02 & 3.89 & 2.35 & 3.76 & 35.80 & 42.66 & 23.83 \\
     & &		245 & 7.15 & 3.19 & 1.84 & 4.74 & 1.99 & 1.15 & 2.96 & 31.29 & 36.92 & 75.24\\
     & 2\%&		92 & 21.28 & 19.66 & 13.43 & 14.19 & 6.37 & 4.35 & 4.59 & 19.81 & 29.17 & 15.83\\
     &	&	171 & 16.88 & 6.55  & 3.96 & 6.33 & 2.12 & 1.28 & 2.05 & 47.28  & 78.18  & 122.38 \\
     & 	&	245 & 7.15 & 3.36  & 1.94 & 4.99 & 1.09 & 0.62 & 1.62  & 139.02 & 67.66 & 137.89 \\
     & 3\% &	92 & 21.28 & 21.97 & 15.01 & 15.86 & 2.67 & 1.83 & 1.91 & 47.63  & 69.73  & 36.86 \\
     & &	171 & 16.88  & 7.32 & 4.43 & 7.08 & 0.88 & 0.53 & 0.85  & 113.27   & 187.31   & 154.35 \\
     &	&	245 & 7.15 & 3.76 & 2.17 & 5.58  & 0.45 & 0.26 & 0.67 & 93.49  & 162.10  & 130.64 \\
\hline
\end{tabular}
\caption{Same as Table~\ref{tab4} but for NGC 5128~halo. Note: All columns have the same meaning as Table~\ref{tab4} but for NGC 5128 galaxy halo. \label{tab7}}
\end{table}
\end{landscape}

\section{Results and~Discussions \label{results}}
The observed Doppler shift \cite{de2011possible, de2014planck, de2015planck, gurzadyan2015planck, de2016triangulum, gurzadyan2018messier, de2019rotating} had been predicted earlier to search for cold gas clouds in Ref.~\cite{de1995observing}, which could contain some of the missing baryons. It has since been used to try to investigate the halo rotation (see Refs.~\cite{tahir2019constraining, tahir2019seeing}). Here, the virial model in Ref.~\cite{qadir2019virial} is used to obtain more reliable and precise answers regarding the halo rotation and the composition and distribution of the clouds in the halos. For~this purpose, we used the following assumptions as a first approximation: (1) all the considered halos contain 1--3\% $f_{dust}$ which remains constant throughout the halo; (2) the entire temperature asymmetry in the CMB $\Delta T/T$ we observed towards the galaxies is due to virial clouds in the halos; (3)~the temperature of halo dust grains is also exactly $2.7$ K; and (4) the virial clouds follow the same dark matter profile NFW, Moore or Burkert models are widely used to map the distribution of dark matter in the galactic~halos. 

The radial distribution of virial clouds  within the M31 halo, taking $f=1$ for definiteness, and~the cumulative radial profiles were obtained by estimating the virial cloud surface density $\Sigma(a)$, by~using Equation~(\ref{d1}), and~then dividing it by the mass of a single cloud (see Figure~\ref{figm31number}). It was seen that the total number of clouds per unit area for each model at large galactocentric distances was approximately the same. We then estimated the cloud-filling factor by using Equation~(\ref{eq5}). Our end goal was to estimate the rotational velocities and the viable chemical composition of the virial clouds of the galaxy halos under consideration for each of the three dark matter models, which are given in Tables~\ref{tab3}--\ref{tab7}. In~most cases, the~obtained values of the rotational velocity appear consistent with the expectations, but~there appeared some inconsistent values in the estimated rotational velocity of the considered halos, which can merely be due to the values chosen for the model parameters, in~particular, $f_{dust}$ and dust temperature at 2.7 K. For~example, at~first, we need to see how dust grains are distributed in the halos of these galaxies and how the halo dust temperature profile can be seen in the halos. The~other point concerns the assumption that the entire temperature asymmetry in the CMB is due to virial clouds. However, four other possibilities may be considered to explain the detected temperature towards the considered galactic halo: (i) the Sunyaev--Zeldovich (SZ) effect, which is the distortion of the CMB through inverse Compton scattering by high energy electrons in galaxy clusters, in~which the low energy CMB photons receive an average energy boost during a collision with the high energy cluster electrons, in~particular, the~kinetic energy \citep{matilla2020probing}; (ii) synchrotron emission by fast-moving electrons which is a type of non-thermal radiation generated by charged particles spiraling around magnetic field lines. Since these electrons are subject to acceleration due to the change of their direction, they emit photons \citep{dolag2000radio}; (iii) free-free emission (or bremsstrahlung emission) by free electrons passing close to atomic nuclei \citep{sun2010galactic}; and (iv) the anomalous microwave emission (AME) from dust grains, which is the electric dipole radiation from small spinning dust grains. The~anomalous component at 53 GHz is 2.5 times as bright as the free-free emission traced by $H\alpha$, providing an approximate normalization for models with significant spinning dust emission \citep{leitch2013discovery}. The spinning dust model is thought to be the simplest explanation for the observed dust correlations, in~the microwave data, but~the combination of increasingly sophisticated models and observations with better frequency resolution will ultimately decide whether spinning dust can explain the detected anomalous emission. So, to analyze the rotational dynamics of galactic halos in more detail by using {\it Planck} data, it would be necessary to constrain the contribution of each of the effects (i)--(iv) above in the temperature asymmetry towards considered halos. In~particular, the~contribution of the rotational kinetic Sunyaev--Zeldovich (rkSZ) has been considered in the case of the M31 halo, and it was found that at~the present level of accuracy, it gives a  negligible contribution to the observed asymmetry towards M31 as seen by the {\it Planck} data~\cite{tahir2022rksz}. Consideration of the contribution of the other effects and the halo dust profile will be attempted in a separate~paper.

We stress that we do not claim that the virial clouds should contain {\it all} the baryonic dark matter present in the galactic halos and should solely explain the observed rotational asymmetry seen towards the considered galaxies in the CMB by {\it Planck} data. Many models that explain the existence of this kind of small dense clouds have been proposed in the~literature.

To take the matter of the true chemical composition and nature of virial clouds more physically, one would need to trace the cloud evolution from the time of their formation to the present, allowing for all the major qualitative changes that occurred during that time. Their formation would have been after the recombination time (that is at $z<1000$), and~they must have been in equilibrium with the CMB since then. The~first step of the evolution of these clouds from the surface of the last scattering (LSS) up to the formation of population-III stars has been studied in detail \citep{tahir2021evolution}.  In~Ref.~\cite{bromm1999primordial} hydrodynamic simulations were used to explain the formation of first-generation stars and investigate the fragmentation process of a metal-free gas. It was seen that this gas collapsed at $z=30$ in clouds with a mass of about $10^6~M_{\odot}$. Fragmentation then led to the formation of dense clumps of mass about $10^3~M_{\odot}$. It was seen in Ref. \citep{tahir2021evolution} that at $z = 50$ virial clouds which are made of primordial gas have a mass of about $10^5~M_{\odot}$ which does not violate the results of Ref.~\cite{bromm1999primordial}. One could take the next step forward and move on towards evolution II of these clouds to understand their true nature, but~the next step of evolution is more problematic and ambiguities can arise. At~this stage, we will have to incorporate the quantum calculations that were found to be negligible in the Evolution I paper. All these things should be taken into account to understand if the virial cloud model used in this paper is not in contradiction with the models already present in literature for such kinds of clouds in the halos. This will be studied later in more detail.
\section*{Acknowledgments}
We acknowledge the use of Planck's data in the Legacy Archive for Microwave Background Data Analysis (LAMBDA) and HEALPix~\cite{gorski2005healpix} packages. The~TAsP and Euclid INFN projects are also acknowledged as well as the anonymous referees for the useful comments.

\end{document}